\documentclass[twocolumn,showpacs,preprintnumbers,amsmath,amssymb,psfig]{revtex4}

\usepackage{graphicx}
\usepackage{dcolumn}
\usepackage{bm}
\def\t{{\rm t}}
\def\K{{\rm K}}
\def\la{\langle}
\def\ra{\rangle}

\newcommand{\beq}{\begin{equation}}
\newcommand{\eeq}{\end{equation}}
\newcommand{\beqa}{\begin{eqnarray}}
\newcommand{\eeqa}{\end{eqnarray}}

\begin{document}
\title{Quantum particle displacement by a moving localized
potential trap}
\author{Er'el Granot}
 \email{erel@yosh.ac.il}
 \affiliation{Department of Electrical and Electronics Engineering, College of Judea and Samaria, Ariel, Israel\\
}
\author{Avi Marchewka}%
 \email{Avi.marchewka@gmail.com}
\affiliation{Department of Electrical and Electronics Engineering, College of Judea and Samaria, Ariel, Israel\\
}
\begin{abstract}
 We describe the dynamics of a bound state of an
attractive $\delta$-well under displacement of the potential.
Exact analytical results are presented for the suddenly moved
potential. Since this is a quantum system, only a fraction of the
initially confined wavefunction remains confined to the moving
potential. However, it is shown that besides the probability to
remain confined to the moving barrier and the probability to
remain in the initial position, there is also a certain
probability for the particle to move at double speed. A
quasi-classical interpretation for this effect is suggested. The
temporal and spectral dynamics of each one of the scenarios is
investigated.
\end{abstract}
 \maketitle

\section{Introduction}
Recent developments in nanotechnology allow displacing miniscule
particles, which can be as small as an atom. These particles'
relocation can be achieved either by optical tweezers
\cite{tweezers} or by Scanning Tunneling Microscopy
(STM)\cite{STM}. The STM moves an atom by creating a potential
well at its vicinity. The atom is then trapped in the tip of the
STM's needle and can easily be relocated along with the tip's
position (see Fig.1). Beautiful structures with incredible (sub
angstrom) accuracy were achieved \cite{STMgallery}.

 Since the atom is a quantum particle,
localization at finite space is always partial. The sudden
activation of the trapping well could cause an atom loss like in
an equivalent decay process \cite{KM05,MG08,MugaWeiSnider}.
Moreover, in this paper we show that the sudden movement itself
(not only the abrupt capturing) can be responsible for the atom
escape. It is also shown that not only do some of the atoms remain
(on the average) at their initial state, but some will move beyond
the tip's influence at \emph{double velocity}.

Bound states subjected to sudden perturbations have been studied
in relation to the so-called deuteron problem \cite{SM05}. The
tunneling dynamics of a bound state has been reported in a time
dependent well \cite{KM05} and after suddenly weakening the
strength of the potential \cite{MG08}. Here we describe the
transport of particles initially trapped in a well which is
shifted at constant velocity along a waveguide. Under strong
transverse confinement, the dynamics becomes effectively
one-dimensional whenever all relevant energies are much smaller
than the excitation quantum in the radial direction.

 \begin{figure}
\includegraphics[width=8cm,bbllx=50bp,bblly=300bp,bburx=500bp,bbury=800bp]{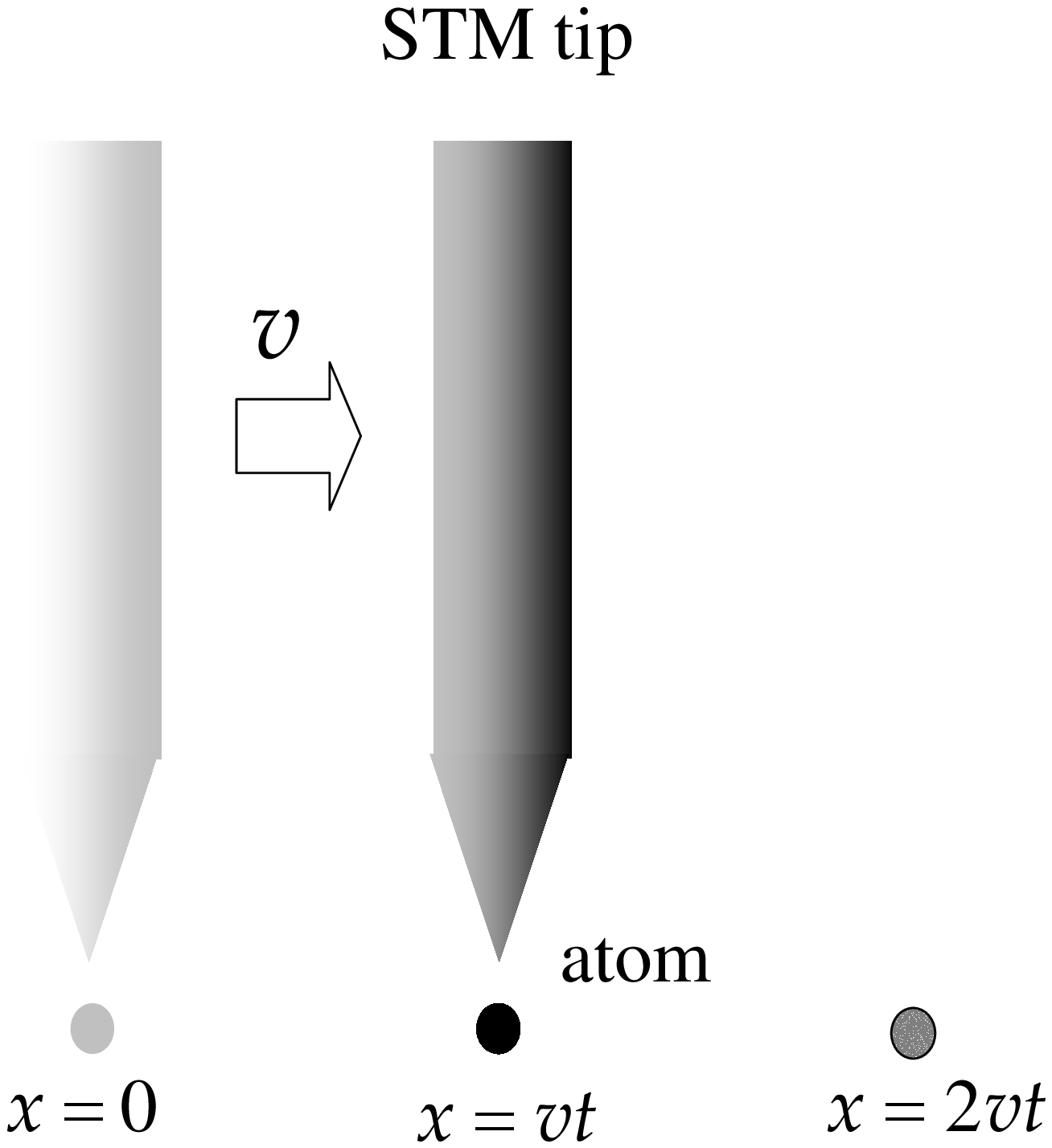}
\caption{System schematic. A single atom displacement by an STM
tip. \label{STMtip}}
 \end{figure}

\section{The model}

To simplify the system, the well is modelled by a one-dimensional
delta function potential well. It should be stressed that a 1D
negative(positive) delta potential is, in fact, an exponentially
shallow potential well, and can model with great accuracy any well
(barrier) whose physical dimensions are smaller than the
de-Broglie wavelength of the particle \cite{Granot_11}.

Prior to $t=0$ the well is localized at $x=0$; however, for $t>0$
the well moves at constant velocity $v$. The potential well can
then be formalized

\begin{equation}
\label{eq1} V\left( {x,t} \right) = \left\{
{{\begin{array}{*{20}c}
 { -\gamma \delta \left( x \right)} \hfill & {\mbox{for}} \hfill & {t \le
0} \hfill \\
 { -\gamma \delta \left( x-vt \right)} \hfill & {\mbox{for}} \hfill & {t >
0} \hfill \\
\end{array} }} \right.
\end{equation}

Thus, the system dynamics are fully characterized by the
Schr\"{o}dinger equation:

\begin{equation}
i\frac{\partial\psi}{\partial t}=-\frac{\partial^2\psi}{\partial
x^2}+V(x,t)\psi
\end{equation}

where we adopted the units $\hbar=2m=1$.

The dynamics begin with the bound state of the $t<0$ potential,
i.e.,

\beqa \label{bound} \psi(x,t=0)=\sqrt{\gamma/2}e^{-\gamma |x|/2}.
\eeqa

\section{Free evolution of the bound state}

If it hadn't been for the potential well, the particle's
probability density would spread out freely. If it is assumed that
for $t>0$ the well is absent, and the Hamiltonian becomes purely
kinetic, then for $t>0$ the dynamics is free, and it can be
obtained using the superposition principle
\beq \psi^{(\infty)}(x,t)=\int_{-\infty}^{\infty}dx'
\K_0(x,t|x,t'=0)\psi(x',t'=0) \eeq
with the free propagator
\beqa \K_0(x,t\vert x',t')=\bigg[\frac{1}{4\pi i
(t-t')}\bigg]^{\frac{1}{2}}e^{i\frac{(x-x')^{2}}{4 (t-t')}}. \eeqa
%
%
%
%
The time evolution of the initial bound state can be then simply
written in terms of the Moshinsky function as
\beqa \label{free}
\psi_0(x,t)&=&\sqrt{\gamma/2}[M\left(x,-i\gamma/2,2t\right)
+M\left(-x,-i\gamma/2,2t\right)].\nonumber\\
\eeqa
where the Moshinsky function reads
\cite{Moshinsky52,Moshinsky76,DMK08}

\beq \label{moshi} M(x,k,\t):=\frac{e^{i\frac{x^{2}}{2
\t}}}{2}w(-z), z=\frac{1+i}{2}\sqrt{\t}\left(k-\frac{x}{\t}\right)
\eeq
in terms of the Faddeyeva function $w(z)$, which is defined as
$w(z):= e^{-z^{2}}{\rm{erfc}}(-i z)$. On physical grounds it is
clear that each of the $M$ functions corresponds to a freely
time-evolved cut-off plane-wave. Such solution entails the
well-known diffraction in time phenomenon, which consists of a set
of oscillations in the density profile
\cite{Moshinsky52,Moshinsky76,DMM07}. However, the imaginary
wavector $-i\gamma/2$ makes such transients evanescent
\cite{MB00}, leading to a uniform expansion.

\section{Uniformly moving well}

 For $t>0$ the propagator should be extended to the $v\neq0$ case, for which the corresponding propagator can be obtained using
Duru's method \cite{Duru89} and as well as by means of the path
integral perturbation series \cite{Grosche93}. It can be
conveniently written in terms of the free propagator
and a perturbation term, represented by a Moshinsky function %
\beqa \label{movingprop}
& & \K_{\delta}^{(v)}(x,t|x',t')= \K_0(x,t|x',t')\nonumber\\
& & +(\gamma/2) e^{i\frac{1}{2}\Big[v(x-vt)-v(x'-vt')
+\frac{v^2(t-t')}{2}\Big]}\nonumber\\
& & \times M\left(|x-vt|+|x'-vt'|,+i\gamma/2,\t\right) \eeqa
%
 \begin{figure}
\includegraphics[width=7.7cm,angle=0,bbllx=40bp,bblly=170bp,bburx=560bp,bbury=600bp]{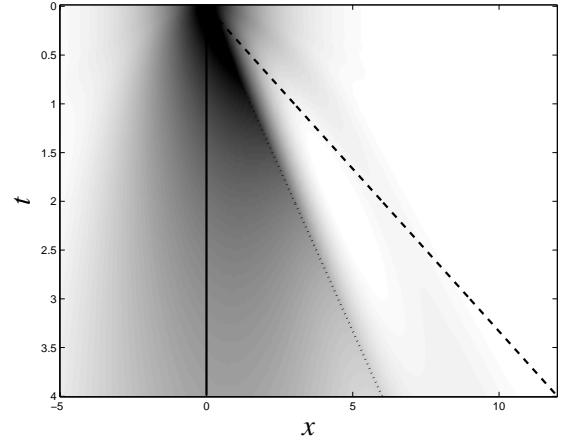}
\caption{Transients developed in the density profile of a bound
state $|\psi(x,t)|^2$, for a moving delta with $v=1.5$ and
$\gamma=0.7$. The dotted line follows the trajectory of the well
$x=vt$, while the dashed line corresponds to $2vt$. Units
$\hbar=2m=1$ are used in all figures. \label{delta}}
 \end{figure}
The time evolution of the state in Eq. (\ref{bound}) can be also
obtained in close-form, using the integral \cite{EK88}
\beqa \label{integral}
& &\int_{-\infty}^{0}dx'e^{ikx'}M(|x'|+|x'|,-iV_0,t)\nonumber\\
&=& \frac{1}{V_0-ik}[M(|x|,k,t)-M(|x|,-iV_0,t)]. \eeqa

Taking $t'=0$ and using Eqs. (\ref{bound}), (\ref{movingprop}) and
(\ref{integral}), one can readily find
\beqa
& & \psi_{\delta}^{(v)}(x,t)=\psi_0(x,t)\nonumber\\
& &
-e^{ivx/2-i\frac{v^2t}{4}}\frac{(\gamma/2)^{\frac{3}{2}}}{\gamma-iv/2}
\nonumber\\
& &\times \big[M(|x-vt|,-v/2-i\gamma/2,2t)-M(|x-vt|,i\gamma/2,2t)\big]\nonumber\\
& &
-e^{ivx/2-i\frac{v^2t}{4}}\frac{(\gamma/2)^{\frac{3}{2}}}{\gamma+iv/2}
\nonumber\\
& &\times \big[
M(|x-vt|,v/2-i\gamma/2,2t)-M(|x-vt|,i\gamma/2,2t)\big],
\nonumber\\
\eeqa
as the sum of a free term plus a perturbation.

%
%
%
The adiabatic Massey parameter \cite{EK88}, which distinguishes
the distinct dynamical regimes is therefore \beqa \theta \equiv
\frac{v}{\gamma}, \eeqa so that for $\theta\ll1$ the adiabatic
dynamics is recovered while $\theta\gg1$ corresponds to the
infinitely fast displacement of the well (free evolution).

Indeed, the eigenstate of a moving delta well is \cite{Duru89}
\beqa
\psi_{b}^{(v)}(x,t)=\sqrt{\gamma/2}e^{i\frac{\gamma^2-v^{2}}{4}t}e^{ivx/2}e^{-\gamma|x-vt|/2},
\eeqa which for $t=0$, becomes $e^{ivx/2}\psi(x,0)$. The fraction
that remains bounded for $t\rightarrow\infty$ is \beqa
|\la\psi_{b}^{(v)}(0)|\psi(0)\ra|^2=\frac{\gamma^4}{(\gamma^2+v^{2})^{2}}=\frac{16}{(4+\theta^{2})^{2}},
\eeqa for $v\gamma\ll1$, the exponential becomes $e^{ivx/2}\sim 1$
in the spatial range of the initial bound state, and the overlap
becomes unity.

\section{Three scenarios}

When $t \rightarrow \infty$ the three domains that were discussed
at the introduction (the particles that remain at the vicinity of
$x=0$, the ones that are localized to the well at uniform velocity
and the ones that propagate at double velocity) eventually appear.
The wavefunction can be written as a supperposition of three
terms:

\beqa \label{three} \psi \sim \psi_{free}+\psi_{well}+\psi_{2v}.
\eeqa where

\beqa
\psi_{free}(x,t)&\cong&\sqrt{\frac{2}{i\pi\gamma t}}e^{i\frac{x^2}{4t}}\frac{1}{1+(x/\gamma t)^2},\nonumber\\
\psi_{well}(x,t)&\cong&\frac{\sqrt{\gamma/2}}{1+(v/2\gamma)^2}e^{i\frac{v}{2}x+i\frac{(\gamma^2-2v^2)t}{4}-\frac{\gamma}{2}|x-vt|},\nonumber\\
\psi_{2v}(x,t)&\cong&\frac{1}{1+iv/2\gamma} \frac{\sqrt{it\gamma/8\pi}}{(x-2vt)+i\gamma t}e^{i\frac{x^2}{4t}-iv^2t}.\nonumber\\
\label{tshort} \eeqa

The first term $\psi_{free}$ describes the free evolution of the
initial state in the absence of the well, as can be appreciated
from the structure of the propagator. The second contribution
remains localized in the moving trap and follows its classical
trajectory $x=vt$, while the last term is responsible for the
appearance of a peak in the density profile at $x=2vt$. This
contribution results from the partial reflection from the
attractive well of the initial state probability density located
at $x>vt$.

Since in the initial state, the particle was localized in a region
as small as $\Delta x \sim \gamma^{-1}$ the uncertainty in the
particle's velocity behave like $\Delta v \sim \gamma$ and
therefore, as can be seen in Eq. \ref{three}, the spatial width of
the two peaks $0$ and $2v$ gets wider approximately like $\Delta x
\sim \gamma t$ (unlike the width of the localized part, which
remains $\sim \gamma^{-1}$). Therefore, the distinction between
the three parts can appear only when $v>\gamma$ (or $\theta>1$).
Moreover, due to their initial width, the peaks shape appears only
when $t \gg (v\gamma)^{-1}$.

The probability density of the exact solution with a comparison to
the approximation, which focuses on the three terms is illustrated
in Fig.3.

 \begin{figure}
\includegraphics[width=7.5cm,angle=0]{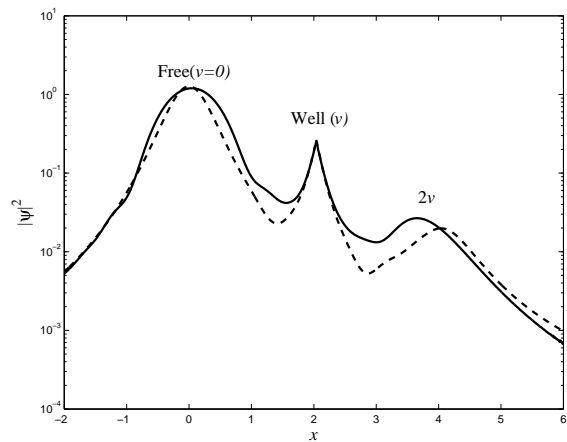}
\caption{Spatial distribution of the probability density
$|\psi|^2$ (solid line) and its approximation (dashed line). The
three domains ($v=0$,$v$, and $2v$) are marked. The parameters are
$\gamma=10$, $v=40$ and $t=0.05$. \label{approx}}
 \end{figure}

\section{Asymptotics}

The transition from the initial stationary bound state to the
final moving one involves the two natural frequencies of the
system: The frequency (energy) of the initial state
$f_1=\gamma^2/8\pi$ and the kinetic energy of the moving particle
$f_2=v^2/8\pi$. The $x=0$ and the $x=2vt$ peaks are affected only
by the frequency $f_2$, however, the $x=vt$ one oscillates with
three harmonics $f_1$, $f_2-f_1$ and $2f_2-f_1$. In Fig.4 the
temporal dynamics of the three peaks is shown. The $x=0$ and the
$x=2vt$ decay like $\sim t^{-1/2}$, while the $x=vt$ one converges
to its final constant value

\begin{equation}
|\psi(x=vt,t\rightarrow\infty)|^2=\frac{\gamma/2}{(1+(v/2\gamma)^2)^2}.
\end{equation}

 \begin{figure}
\includegraphics[width=7.5cm,bbllx=100bp,bblly=430bp,bburx=480bp,bbury=780bp]{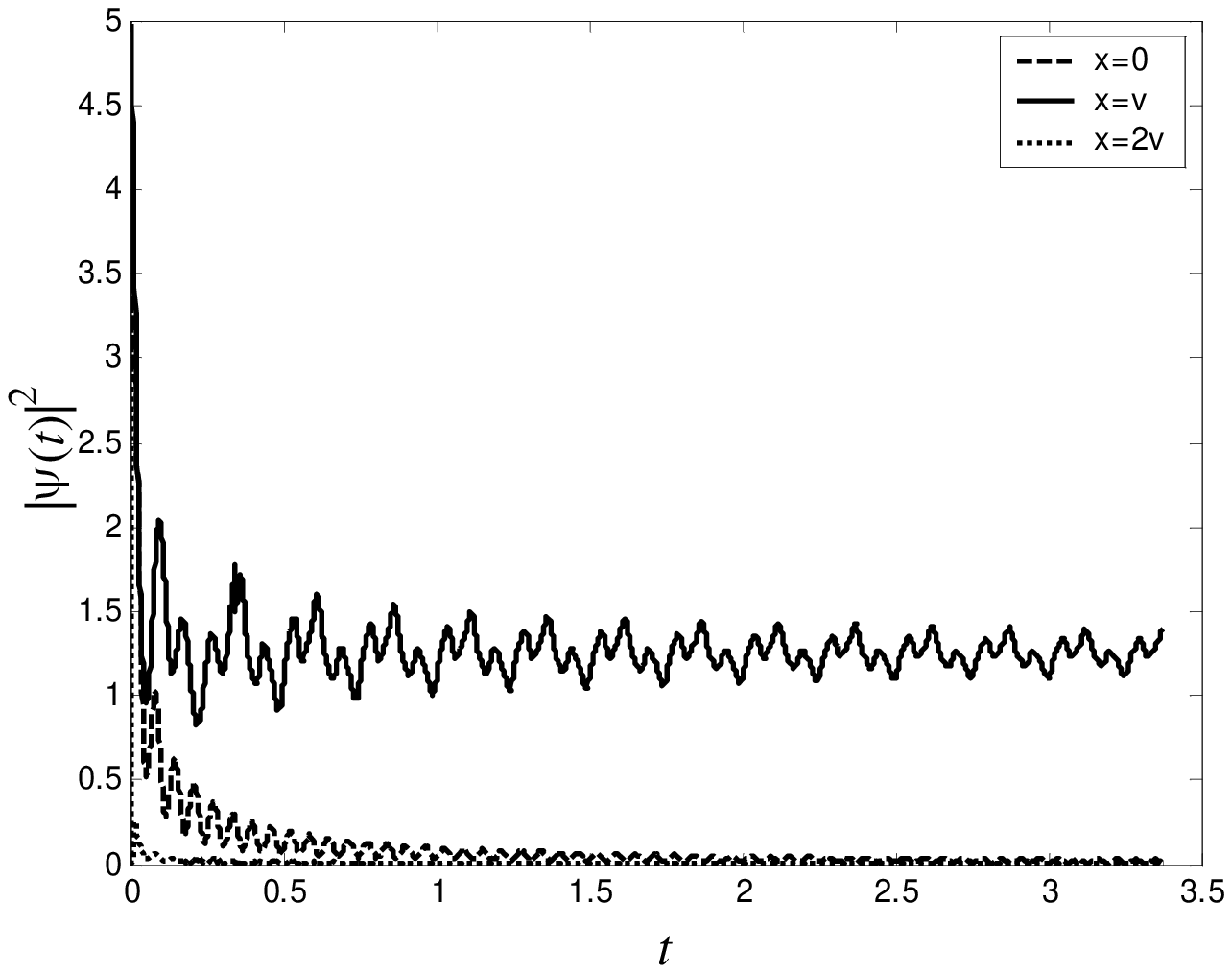}
\caption{The temporal dynamics of the three peaks ($x=0$, $x=vt$
and $x=2vt$). The system's parameters are: $\gamma=10$, $v=20$.
\label{temp_peak2}}
 \end{figure}

In Fig. 5 a numerical spectral distribution $\Psi(f) \equiv
FFT[|\psi|^2]$ (FFT stands for the Fast Fourier Transform) of each
one of the peaks is presented. The four different frequencies are
clearly shown.

The 'A' peak corresponds to the frequency $f_1=\gamma^2/8\pi$. The
'D' and 'E' peaks correspond to the frequency $f_2=v^2/8\pi$, the
'B' one stands for $f_3=|f_2-f_1|=|v^2-\gamma^2|/8\pi$ and finally
'C' stands for $f_4=f_3+f_2=|2v^2-\gamma^2|/8\pi$.

 \begin{figure}
\includegraphics[width=7.5cm,bbllx=100bp,bblly=430bp,bburx=480bp,bbury=780bp]{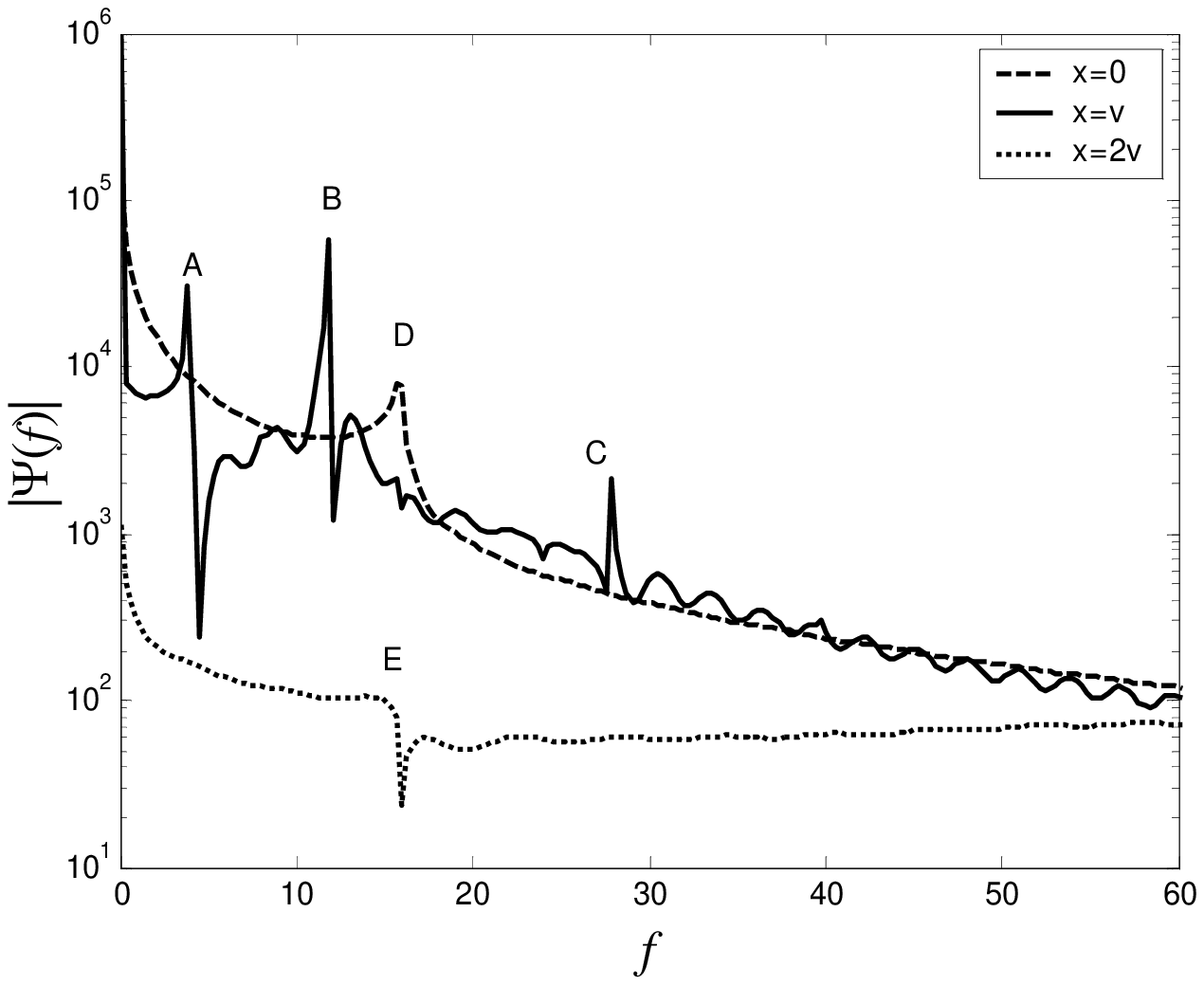}
\caption{The spectral distribution $\Psi(f)$ of the three peaks
(of Fig.4). \label{spect2}}
 \end{figure}

\section{Semi-classical realization}
Clearly, the double velocity effect cannot be classical, since a
localized state will remain localized in the classical world.
However, the origin of this effect has partially a classical
interpretation. Quantum mechanically, the particle in the initial
state is not completely localized inside the well. In fact, when
the well is very narrow most of the chance is to find the
localized particle outside the well.

It is also instructive to investigate the system in a moving frame
of reference, in which the well is at rest (originally, at the lab
reference the well moves to the right).

At $t=0$ the particle (at the well's reference frame) begins to
move to the left with respect to the well. We can regard it as
three different scenarios: particle at the right of the well
(gray) at the well (black) at its left (white) At $t=0$ all three
types begin to moves simultaneously to the left at velocity $v$
(Fig.6A).

The white is free - so it remains at velocity v to the left. The
black is trapped - so its average velocity is zero. But the gray
hits the barrier and turns back with velocity $-v$, i.e., the
final scenario is shown at Fig.6B.

When we return to the lab frame of reference (where the well moves
to the right), we see that the white one didn't move, the black
moved with the well at velocity $v$ and the gray moved with
velocity $2v$ (Fig.6C).

Obviously, this semiclassical interpretation is possible only due
to the partial localization, which is a manifestation of Quantum
mechanics.

 \begin{figure}[h!]
\includegraphics[width=7.5cm,bbllx=70bp,bblly=400bp,bburx=350bp,bbury=800bp]{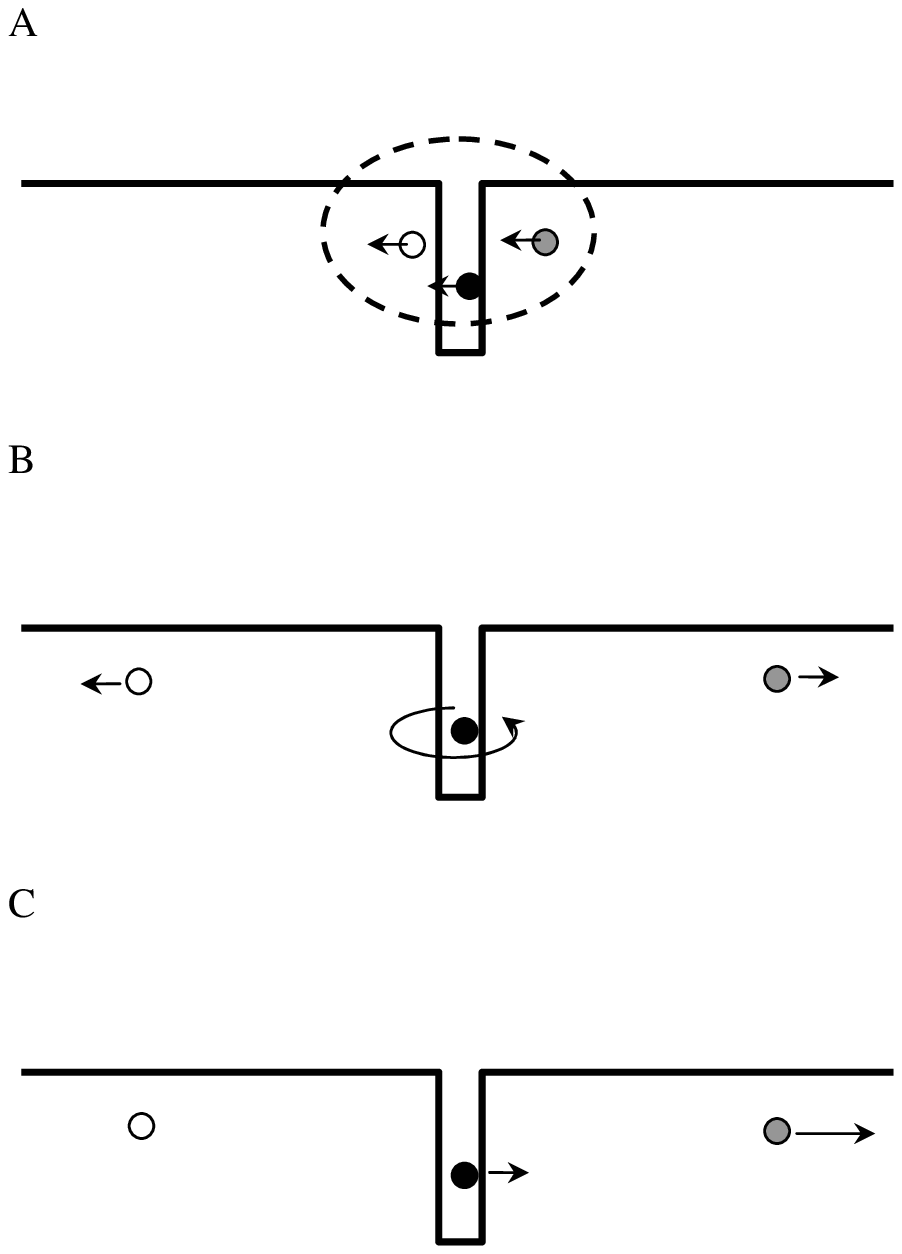}
\caption{Semiclassical realization of the three domains.
\label{classical}}
 \end{figure}

\section{Conclusion and Discussion}

We have presented a 1D quantum model for an atom displacement with
an STM tip. The model consists of a delta function well, which
model the STM's potential at the vicinity of its tip end, which
moves uniformly. It was shown that the probability to remain
trapped in the moving tip is
$\frac{\gamma^4}{(\gamma^2+v^{2})^{2}}=\frac{16}{(4+\theta^{2})^{2}}$.

Moreover, it was shown that besides the trapped particles, and the
particles that remain close to their initial state, there is also
a third group of particles, which propagates at double velocity
($2v$) away from their initial position. We show that the
probability for each one of the three groups has a different
temporal dynamics.

\begin{acknowledgments}
Our appreciation goes to A. del Campo for providing the early
motivation for the model and for our subsequent discussions.
\end{acknowledgments}

\end{document}